\documentclass[aps,prl,twocolumn,superscriptaddress,showpacs,showkeys]{revtex4}

\usepackage{graphicx}
\usepackage{amsmath}

\begin{document}


\title{Broken Time Reversal and Parity Symmetries for Electromagnetic\\
Excitations in Planar Chiral Nanostructures}


\author{Ewan M. Wright}
\email{ewan.wright@optics.Arizona.EDU} \affiliation{Optical
Sciences Center and Department of Physics, University of Arizona,
Tucson, AZ 85721, USA}

\author{Nikolay I. Zheludev}
\email{n.i.zheludev@soton.ac.uk}
\affiliation{School of Physics and Astronomy, University of
Southampton, SO17 1BJ, UK}


\date{\today}

\begin{abstract}
We elucidate the physical mechanisms by which electromagnetic
excitations in planar chiral waveguides exhibit broken time
reversal and parity symmetries as recently observed in metallic
nanostructures ({\sl Phys. Rev. Lett.} {\bf 91}, 247404 (2003)).
Furthermore, we show that concomitant with these broken symmetries
the electromagnetic excitations can acquire fractional winding
numbers, revealing analogies between light scattering from planar
nanostructures and anyon matter.
\end{abstract}
\maketitle
%
A recent experimental study of optical interactions with
non-magnetic metal planar chiral nanostructures consisting of
arrays of four-fold chiral gammadions provides intriguing evidence
of broken time reversal (${\cal T}$) symmetry \cite{planar,
archive}. Planar chiral objects are characterized by the fact that
they cannot be brought into congruence with their enantiomeric or
mirror image form without lifting the object from the plane. Among
the intriguing properties of chiral arrays is that they display
non-reciprocal polarization effects in diffraction which resemble
the well-known non-reciprocity of the Faraday effect
\cite{archive2}. In the experiment of Ref. \cite{planar} chiral
nanostructures whose topographies were mirror images of each other
were found to lose there mirror symmetry when viewed in polarized
light, an effect that also occurred for individual chiral
gammadions. It has been proposed that these unusual symmetry
properties result from the chiral optical response associated with
plasmon and volume modes in the grooves of the structure
\cite{planar}.

In this paper we theoretically show that electromagnetic (EM)
excitations in planar chiral nanostructures can exhibit broken
${\cal T}$ and parity (${\cal P}$) symmetries, and a physical
argument is given for how these broken symmetries can be
manifested in the properties of light scattered from the
nanostructures. We also point to analogies between light
scattering from planar chiral nanostructures and from anyon matter
\cite{Rojo92}.

\begin{figure}
\includegraphics[width=80mm]{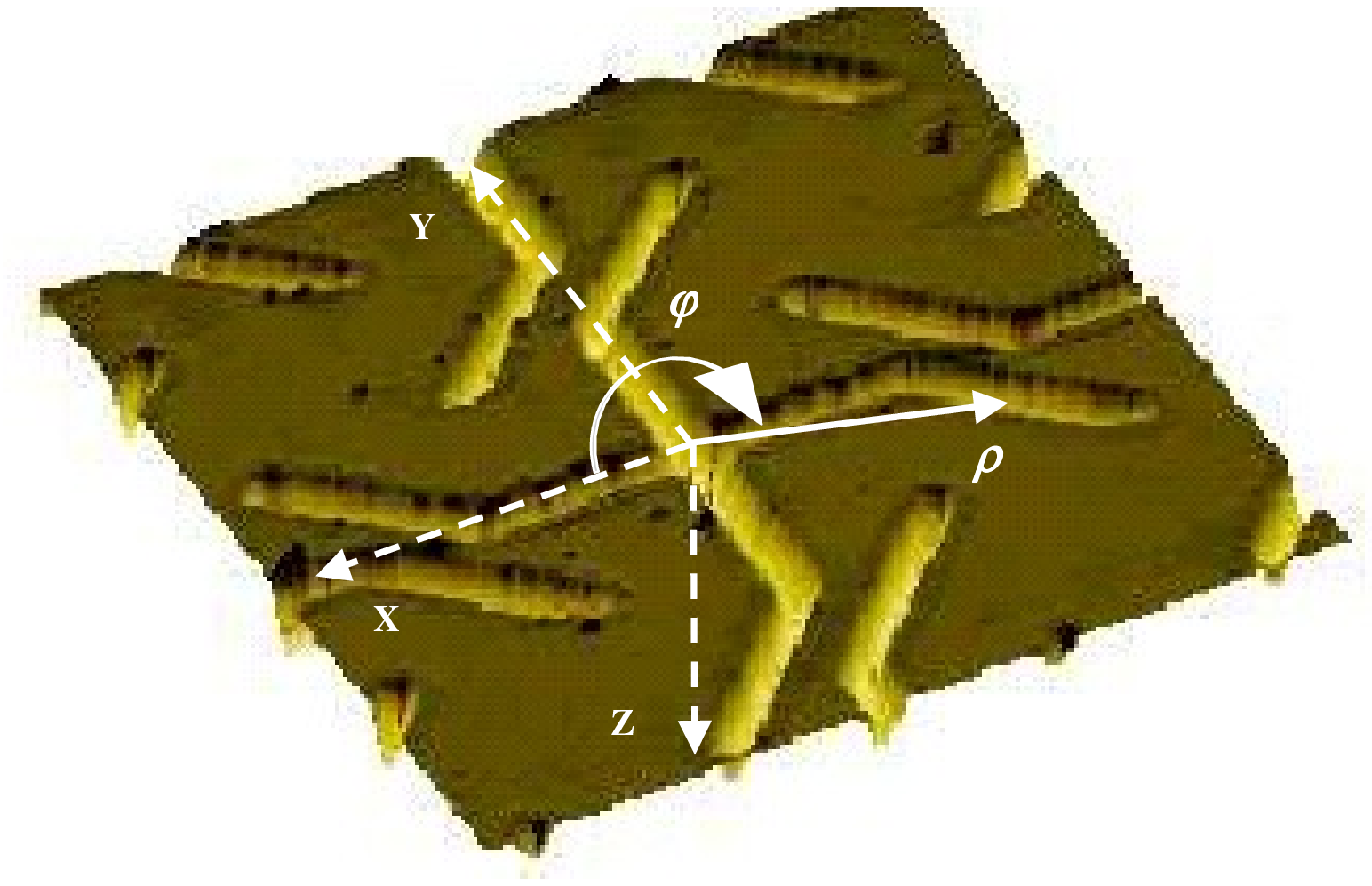}
\caption{[colors online] An AFM image of a fragment of the planar
chiral structure showing one gammadion carved in a thin
gold/titanium film, as described in Ref. \cite{archive}. The
electromagnetic excitations considered in the present paper are
concentrated in the grooves of the structure that form a single
gammadion element. The characteristic sizes of the gammadion's
elements, their width and depth are 1.5 $\mu$m, 700 nm, and 120
nm, respectively.} \label{anyon}
\end{figure}

To describe the experiment of Ref. \cite{planar} we consider a
thin metallic (gold-titanium) film with an array of four-fold
chiral gammadions carved in it. We neglect coupling between
individual gammadions based on the high losses suffered by
propagating surface plasmon-polariton waves propagating between
gammadions due to the titanium layer covering the structure. These
losses, however, are negligible for EM excitations confined in the
grooves of individual gammadions as considered here, since the
titanium layer is absent in the grooves. We thus consider an
individual four-fold chiral gammadion centered on the origin, an
example of which is shown in Fig. \ref{anyon}. Here the gammadion
lies in the $(x,y)$ plane and we have chosen to align the x-axis
with the central portion of one gammadion groove and the y-axis
with another, the z-axis being perpendicular to the plane of the
gammadion. For future use we also introduce a cylindrical
coordinate system for which points in the plane of the gammadion
are represented by the radial coordinate $\rho$ and azimuthal
angle $\phi$ as indicated in Fig. \ref{anyon}. Then, for an EM
field of frequency $\omega$ the dielectric tensor around the
gammadion structure has elements $\epsilon_{\mu\nu}(\vec{r})$,
with $\mu,\nu=x,y,z$, where we omit the explicit $\omega$
dependence for simplicity in notation. We approximate the tensor
as Hermitian and its diagonal elements as real by the virtue of
the fact that for the optical frequencies of interest in gold
$\mid Re{(\epsilon_{\mu\mu})}\mid \gg \mid
Im{(\epsilon_{\mu\mu})}\mid$, and the four-fold gammadion symmetry
implies $\epsilon_{xx}=\epsilon_{yy}$. The generally complex
off-diagonal tensor element $\epsilon_{xy}=\epsilon_{yx}^*$
describes polarization coupling in the $(x,y)$ plane due to the
chiral gammadion patterning, and we assume that this is the
dominant source of polarization coupling and hereafter set all
other off-diagonal tensor elements to zero. To proceed we expand
the tensor elements $\epsilon_{xy}(\vec{r})$ into azimuthal basis
functions
\begin{equation}
\epsilon_{xy}(\vec{r}) = \sum_{\ell=-\infty}^{\infty}
\epsilon_{xy}^{(\ell)}(\rho,z)e^{i\ell\phi} . \label{expand}
\end{equation}
%
%
%
%
%
The basis functions $\exp(i\ell\phi)$ are useful for
characterizing the chirality of the structure as each has an
associated integer winding number $\ell$ such that the phase
varies from zero to $2\pi\ell$ upon circling around the z-axis.
The reality of the diagonal dielectric tensor elements implies
that they have equal weighting of the winding numbers $\pm\ell$,
$\epsilon_{\mu\mu}^{(\ell)}=(\epsilon^{(-\ell)}_{\mu\mu})^*$,
meaning that the chirality is not manifest in the diagonal
elements. In contrast, the generally complex off-diagonal tensor
elements can manifest the chirality of the gammadion. For
illustrative purposes we consider the simple case that the
off-diagonal elements are dominated by a single winding number
$\ell$, the contribution from $-\ell$ being set to zero, for
example $\ell=4$ for a gammadion with four-fold symmetry and
$\ell=-4$ for the corresponding enantiomeric form, and we write
the off-diagonal elements as
$\epsilon_{xy}(\vec{r})=(\epsilon_{yx}(\vec{r}))^*=\kappa(\rho,z)e^{i\ell\phi}$.
Finally, for the diagonal elements we ignore, to lowest-order, the
patterning of the gammadion, so that the diagonal elements
$\epsilon_{\mu\mu}(z)$ are functions of $z$ only, so the
dielectric tensor becomes
\begin{equation}
\epsilon(\vec{r}) = \left ( \begin{array}{ccc} \epsilon_{xx}(z) &
\kappa(\rho,z)e^{i\ell\phi} & 0
\\ \kappa(\rho,z)e^{-i\ell\phi} & \epsilon_{xx}(z) &
0\\ 0 & 0 & \epsilon_{zz}(z)
\end{array} \right ) . \label{diel}
\end{equation}
The dielectric tensor (\ref{diel}) describes essentially a
non-local medium for which the off-diagonal tensor elements are
space-dependent through the chiral terms $\exp(\pm i\ell\phi)$,
and we have taken the function $\kappa(\rho,z)$ to be real so as
to highlight the effects of the chirality. We further assume that
$|\epsilon_{\mu\mu}|>>|\kappa|$.

To proceed we examine EM excitations that are bound to an
individual chiral gammadion or waveguide. These EM waves are
time-harmonic solutions of the Maxwell equations in the presence
of the gammadion waveguide with electric fields of the form
$\vec{E}(\vec{r},t)=[\vec{E}(\vec{r},\omega,t)e^{-i\omega t}
+c.c.]/2$. In the slowly-varying envelope approximation when the
vector electric field envelope $\vec{E}(\vec{r},\omega,t)$ changes
slowly on the time-scale $1/\omega$ and we neglect second-order
time derivatives, Maxwell's equations lead to
\begin{equation}
-\nabla\times\nabla\times\vec{E}
+2i\omega\mu_0\epsilon(\vec{r}){\partial\vec{E}\over
\partial t} + \mu_0\omega^2\epsilon(\vec{r})\vec{E}(\vec{r}) =
0 , \label{Max}
\end{equation}
with the generally complex dielectric tensor $\epsilon(\vec{r})$
given by Eq. (\ref{diel}). For transverse electric (TE) waves
whose polarization lies in the $(x,y)$ plane we find, using the
dielectric tensor (\ref{diel}) and neglecting the off-diagonal
elements to dominant order, that
$\nabla\cdot\vec{D}=\nabla\cdot(\epsilon(\vec{r})\vec{E})=\epsilon_{xx}(z)\nabla\vec{E}=0$,
which yields $\nabla\cdot\vec{E}=0$. This restriction to TE
polarization is justified if we assume that the diagonal
dielectric tensor element $\epsilon_{xx}(z)$ provides sufficient
confinement along the z-axis so that the usual designation of TE
polarization (electric field perpendicular to z) and transverse
magnetic (TM) polarization (magnetic field perpendicular to z) for
waveguides applies \cite{Marcuse}. For strong confinement along
the z-axis we assume that the field structure along the
z-direction is dominated by the normalized fundamental mode $u(z)$
determined from
$[d^2/dz^2+\mu_0\omega^2\epsilon_{xx}(z)]u(z)=\beta^2u(z)$, with
$\beta^2$ the corresponding eigenvalue. Then substituting
$\vec{E}(\vec{r},\omega,t)=u(z)[\vec{e}_x{\cal E}_x(\rho,z,t)+
\vec{e}_y{\cal E}_y(\rho,z,t)]$ in Eq. (\ref{Max}), with
$\vec{e}_{x,y}$ denoting unit vectors along the respective
directions, projecting out the mode structure along z, and
introducing the spinor notation $\bar{\cal
E}(\rho,\phi,t)=col\left({\cal E}_x(\rho,\phi,t),{\cal
E}_y(\rho,\phi,t)\right )$, Eq. (\ref{Max}) reduces to the
following effective planar equations for the EM excitations
coupled to the chiral waveguide
\begin{equation}
2i\left ({\beta^2\over\omega}\right ){\partial\bar{\cal E}\over
\partial t} = \hat H\bar{\cal E} , \label{spinor}
\end{equation}
where the Hermitian Hamiltonian is given by
\begin{equation}
\hat H=\hat H_{D}-\left (\begin{array}{cc} 0 &
\mu_0\omega^2\bar\kappa(\rho)e^{i\ell\phi}
\\ \mu_0\omega^2\bar\kappa(\rho)e^{-i\ell\phi}  & 0
\end{array} \right ) , \label{Hell}
\end{equation}
with diagonal part $\hat H_D=-(\nabla_\perp^2+\beta^2)I$, $I$
being the unit $(2\times 2)$ matrix and $\nabla_\perp^2$ the
transverse Laplacian acting in the $(x,y)$ plane, and
$\bar\kappa(\rho)=\int dz |u(z)|^2\kappa(\rho,z)$. We remark that
by projecting out the mode structure $u(z)$ along $z$ we have
rendered the system effectively two-dimensional (2D). However, the
physical system as described by Eqs. (\ref{spinor}) is still
sensitive to the selection of the direction of the z-axis since
the gammadion winding number $\ell$ is defined with respect to the
angular direction of phase advance around the z-axis of the chiral
term $\exp(i\ell\phi)$. The key ingredient for the following
discussion is that we are considering TE waves for which the
electric field has only two vector components instead of three,
and this allows for our two component spinor approach. This 2D
planar approximation for TE waves is valid as long as the
structure is tightly confined and single-mode along the z-axis. We
note that Eqs. (\ref{spinor}) are analogous to the Bogoliubov
equations for electrons in a superconductor \cite{deGennes}, with
$\kappa(\rho,z)$ playing the role of the pair potential.

We next examine the time-reversal properties of the spinor wave
Eq. (\ref{spinor}). For our system a suitable antilinear
time-reversal operator is ${\cal T}=K_0I$, where $K_0$ is the
complex conjugation operator \cite{Ballen}. By operating ${\cal
T}$ on Eq. (\ref{spinor}) we obtain the time-reversed equation
\begin{equation}
2i\left ({\beta^2\over\omega}\right ){\partial\over\partial
(-t)}({\cal T}\bar{\cal E}) = {\cal T}\hat H {\cal T}^{-1}
(\cal{T}\bar{\cal E}) ,
\end{equation}
and for $\hat H$ given in Eq. (\ref{Hell}) we find
\begin{equation}
{\cal T}\hat H {\cal T}^{-1}=\hat H^* \equiv {\cal P}\hat H {\cal
P}^{-1} . \label{Hconj}
\end{equation}
For our 2D reduced system described by the spinor Eq.
(\ref{spinor}) the linear parity operator ${\cal P}$ has the
effect of reversing $\phi\rightarrow -\phi$, which can be
implemented for the Hamiltonian (\ref{Hell}) using ${\cal
P}=\sigma_x$, with $\sigma_\mu$ being the Pauli-spin matrices.
Since a structure and its corresponding enantiomeric form are
interconverted by space inversion \cite{Barron}, Eq. (\ref{Hconj})
tells us that time-reversal is equivalent to spatial inversion.
This feature was noted in Ref. \cite{planar} and simply means that
under time-reversal incident EM waves approach the planar
structure from the opposite direction and perceive it as the
enantiomeric structure of opposite chirality. In the absence of
chirality $\ell=0$, when the Hamiltonian in Eq. (\ref{Hell}) is
purely real, the time-reversed and space inverted Hamiltonians in
Eq. (\ref{Hconj}) are both equal to $\hat H$ meaning that both
${\cal T}$ and ${\cal P}$ symmetries are intact. In contrast, in
the presence of chirality ${\cal T}$ and ${\cal P}$ symmetries are
simultaneously broken for the planar chiral structure. However,
Eq. (\ref{Hconj}) implies that $({\cal PT})\hat H ({\cal
PT})^{-1}=\hat H$, so that the system is always invariant under
the combined action of ${\cal PT}$ symmetry \cite{Rojo92}.

More generally, the off-diagonal matrix element
$\epsilon_{xy}(\vec{r})$ of the dielectric tensor may expressed as
a sum over winding numbers $\ell$ as in Eq. (\ref{expand}), a
special example of which is the single dominant winding number
case considered above. Then, unless
$\epsilon^{(\ell)}_{xy}=(\epsilon^{(-\ell)}_{xy})^*$ for all
$\ell$ and $\epsilon(\vec{r})$ is real, both ${\cal T}$ and ${\cal
P}$ symmetries are simultaneously broken. Essentially the time
reversal and parity asymmetries in planar chiral structures are
underpinned by the nonlocality of the gammadion response, meaning
that excitation of the extended gammadion as a whole is crucial. A
planar structure consisting of individually radiating atoms could
not produce such a symmetry breaking interaction.

We are now in a position to discuss how these new results relate
to the experiment of Ref. \cite{planar} which provided strong
experimental evidence of broken ${\cal T}$ and ${\cal P}$
symmetries. Distinct polarization phenomena for media where ${\cal
T}$ and ${\cal P}$ symmetries are simultaneously broken have
previously been discussed in relation to anyon matter in
high-temperature superconductors, initially by Wen and Zee
\cite{Wen89} and later by Canright and Rojo \cite{Rojo92}. For the
purpose of analyzing how the broken symmetries applicable to TE
excitations may lead to observable effects, it is important to
realize that an incident EM field can couple energy into the
planar TE waves of individual chiral waveguides as strict momentum
conservation is relaxed at the edges of the grooves of the
gammadions, and by the same argument EM energy can be out-coupled
from individual chiral waveguides. Furthermore, previously the
in-plane patterning of the diagonal dielectric tensor element
$\epsilon_{xx}(z)$ was neglected. At next order the radial
dependence of $\epsilon_{xx}(\rho,\phi,z)$ provides a grating
structure that can in-couple and out-couple energy to radiation
modes analogous to the emission mechanism of surface emitting
circular grating lasers \cite{ErdHal90}. This coupling between the
incident, out-coupled, and TE waves of the chiral waveguides,
alongside the losses present in real systems, give the TE waves a
finite lifetime. Furthermore, these mechanisms offer an
explanation of why the fields scattered from chiral planar
nanostructures can manifest the broken ${\cal T}$ and ${\cal P}$
symmetries applicable to the bound TE waves, since the diffracted
fields from the nanostructures have their origin in energy
out-coupled from the TE waves excited by the incident field. Thus,
it is physically reasonable that the scattered EM fields will
inherit the broken ${\cal T}$ and ${\cal P}$ symmetries exhibited
by the planar TE waves.

We have also found that concomitantly with the breaking of the
${\cal T}$ and ${\cal P}$ symmetries the corresponding TE
excitations bound to chiral waveguides can acquire fractional
winding numbers and we now demonstrate this interesting property
which also elucidates the vector nature of the TE waves.
Inspection shows that consistent stationary solutions of Eq.
(\ref{spinor}), corresponding to time-harmonic solutions at
frequency $\omega$, assume the form
\begin{equation}
\bar{\cal E}(\rho,\phi,t)=\left(\begin{array}{c} a_x(\rho)e^{i(m+\ell)\phi}\\
a_y(\rho)e^{im\phi}\end{array}\right ) , \label{SState}
\end{equation}
where the winding number of the x-polarized component of the field
is $(m+\ell)$, and $m$ is that for the y-polarized component,
$m=0,\pm 1,\pm 2,\ldots$ being an integer that we use to label the
solutions. This form of solution ensures consistency between the
azimuthal variation of the electric field polarization components
in the presence of the phase dependent coupling terms $\exp(\pm
i\ell\phi)$ in the dielectric tensor. We introduce the average
winding number for a TE vector field as
\begin{equation}
<W_{EM}> = \frac{\int_0^{2\pi}d\phi\int_0^\infty \rho d\rho
\bar{\cal E}^\dagger(\rho,\phi,t) \hat W_{EM}\bar{\cal
E}(\rho,\phi,t) } { \int_0^{2\pi}d\phi\int_0^\infty \rho d\rho
\bar{\cal E}^\dagger(\rho,\phi,t)\bar{\cal E}(\rho,\phi,t) } ,
\end{equation}
where we have defined the EM winding number operator
$\hat{W}_{EM}= -i(\partial/\partial\phi)I$. For the EM wave
solutions in Eq. (\ref{SState}) we find
\begin{equation}
<W_{EM}> = m +\ell(1+\bar s_1)/2 , \label{WEM}
\end{equation}
where
\begin{equation}
\bar s_1 = \frac{\int_0^{2\pi}d\phi\int_0^\infty \rho d\rho \left
[|a_x(\rho)|^2-|a_y(\rho)|^2\right ] }
{\int_0^{2\pi}d\phi\int_0^\infty \rho d\rho \left
[|a_x(\rho)|^2+|a_y(\rho)|^2\right ]} .
\end{equation}
is the spatially averaged and normalized value of the element
$s_1=(|{\cal E}_x|^2-|{\cal E}_y|^2)$ of the Stokes vector
$\vec{s}=(s_1,s_2,s_3)$ \cite{book2} over the plane of the
individual waveguide. In general, $-1<\bar s_1<1$, so that the EM
winding number in Eq. (\ref{WEM}) can be non-integer, that is,
fractional. The average winding number of the field is associated
with its orbital angular momentum that has its origin in the
spatial structure of the fields \cite{Allen}, as distinct from the
spin angular momentum associated with the polarization state of
the field. The average spin angular momentum is proportional to
the spatially averaged value of the $s_3=2Im({\cal
E}_x(\rho,\phi,t)^*{\cal E}_y(\rho,\phi,t))$ component of the
Stokes vector \cite{book2}, which for $\ell\ne 0$ yields $\bar
s_3=0$ for the EM solutions in Eq. (\ref{SState}) by virtue of the
azimuthal integrals over $\phi$. This implies that in the presence
of planar chirality the state of polarization of the vector field
components varies over the plane in such a way that the average
spin angular momentum is zero, and the average angular momentum of
the coupled fields is equal to the orbital angular momentum
associated with the fractional modal winding number. To see the
consequence of this we realize that here we are discussing the
classical TE vector modes of the structure, and we anticipate that
upon quantization of the field we can add photons to each of these
vector modes \cite{BerSto03}. Associated with each photon added to
vector TE mode of index $m$ there will be an average angular
momentum directed along the z-axis
$<J_z>=\hbar<W_{EM}>=\hbar(m+\ell(1+\bar s_1)/2)$, that is, the
average angular momentum of the quantized vector modes can be
fractional. The average winding number for the EM excitation can
be fractional since the deviation from the integer values expected
for a bosonic field is fully accounted for by the angular momentum
associated with the chiral waveguide \cite{Wil82,JacRed83}.

It is worthwhile commenting on the nature of the EM mode
interaction with the chiral gammadion. As noted earlier,
polarization is associated with the spin of the EM field, whereas
the spatial variation of the EM field describes orbital angular
momentum, for example, a Laguerre-Gaussian laser field
\cite{Allen}. From Eq. (\ref{SState}) we see that for $\ell \neq
0$ the relative phase between the two vector field components
varies with azimuthal angle $\phi$, meaning that the polarization
state of the field is not spatially homogeneous, as is obvious
from the polarized images provided in Ref. \cite{planar}.
Furthermore, since the two vector field components have different
winding numbers they carry different orbital angular momenta.
Thus, the chiral waveguide intertwines the spin and orbital
angular momenta for the EM modes in such a way that they cannot be
separated, in contrast to optical activity, for example, in which
the molecular chirality causes polarization rotation but otherwise
leaves the orbital angular momentum unchanged.

In summary, we have elucidated the physical mechanisms by which EM
waves coupled to chiral waveguides can display broken ${\cal T}$
and ${\cal P}$ symmetries. Symmetry is recovered in the form of
enantiomeric reversibility ${\cal PT}$ based on the equality of
excitations in the forward and time-reversed processes involving
left- and right-handed structures, respectively. We have argued
that these broken symmetries should also be manifest in the fields
scattered from the structures as reported in Ref. \cite{planar}.
Furthermore, we have shown that the TE excitations bound to chiral
waveguides can acquire fractional winding numbers. We finish by
drawing attention to the analogy between the light scattering from
anyon matter and chiral nanostructures, since both involve broken
${\cal T}$ and ${\cal P}$ symmetries and excitations with
fractional angular momentum \cite{Wil82,book}.
\begin{acknowledgments}
The authors thank A. Schwanecke for providing an AFM image of the
planar structure used in Ref. \cite{planar} and fruitful
discussions and acknowledge the support of the Science and
Engineering Research Council (UK).
\end{acknowledgments}

\end{document}